\documentclass[ aps, showpacs, showkeys, nofootinbib, floatfix, superscriptaddress]{revtex4}

\usepackage{amsfonts}

\usepackage{amssymb}
\usepackage{amsmath}
\usepackage{graphicx}

\begin{document}

\title{Cosmic constraint on the  unified model of dark sectors  with or without a cosmic string fluid in the varying gravitational constant theory}

 \author{Jianbo Lu}
 \email{lvjianbo819@163.com}
 \affiliation{Department of Physics, Liaoning Normal University, Dalian 116029, P. R. China}
 \author{Yanfeng Xu}
 \affiliation{Department of Physics, Liaoning Normal University, Dalian 116029, P. R. China}
 \author{Yabo Wu}
 \affiliation{Department of Physics, Liaoning Normal University, Dalian 116029, P. R. China}

\begin{abstract}

  Observations indicate that  most of the universal matter are invisible and the gravitational constant $G(t)$ maybe depends on the time. A theory of the variational $G$ (VG) is explored in this paper, with naturally producing the useful dark components in universe. We utilize the observational data: lookback time  data,  model-independent gamma ray bursts, growth function  of matter linear perturbations, type Ia supernovae data with  systematic  errors,  CMB and BAO  to restrict the unified model (UM) of dark  components in VG theory.  Using the best-fit values of parameters with the covariance matrix,  constraints on the variation of $G$ are $(\frac{G}{G_{0}})_{z=3.5}\simeq 1.0015^{+0.0071}_{-0.0075}$ and $(\frac{\dot{G}}{G})_{today}\simeq -0.7252^{+2.3645}_{-2.3645}\times 10^{-13} yr^{-1}$, the small uncertainties around constants. Limit on the equation of state of dark matter is $w_{0dm}=0.0072^{+0.0170}_{-0.0170}$   with assuming $w_{0de}=-1$  in unified model, and  dark energy is  $w_{0de}=-0.9986^{+0.0011}_{-0.0011}$   with assuming $w_{0dm}=0$ at prior. Restriction on  UM parameters are $B_{s}=0.7442^{+0.0137+0.0262}_{-0.0132-0.0292}$ and $\alpha=0.0002^{+0.0206+0.0441}_{-0.0209-0.0422}$ with  $1\sigma$ and  $2\sigma$ confidence level. In addition,  the effect of a cosmic string fluid on unified model in VG theory are investigated. In this case it is found that the $\Lambda$CDM ($\Omega_{s}=0$, $\beta=0$ and $\alpha=0$) is included in this VG-UM model  at $1\sigma$ confidence level, and the larger errors are given: $\Omega_{s}=-0.0106^{+0.0312+0.0582}_{-0.0305-0.0509}$ (dimensionless energy density of cosmic string), $(\frac{G}{G_{0}})_{z=3.5}\simeq 1.0008^{+0.0620}_{-0.0584}$  and  $(\frac{\dot{G}}{G})_{today}\simeq -0.3496^{+26.3135}_{-26.3135}\times 10^{-13}yr^{-1}$.

\end{abstract}



\maketitle

\section{$\text{Introduction}$}

 { Gravity theories are usually studied with an assumption  that  Newton gravity constant $G$ is a constant. But some observations hint that $G$ maybe depends on the time \cite{VG-t}, such as observations from  white dwarf star  \cite{VG-MNRAS-2004-dwarf,VG-PRD-2004-white},  pulsar  \cite{VG-APJ}, supernovae \cite{VG-PRD-2002-SN} and neutron star \cite{VG-PRL-1996-neutron}. In addition, cosmic observations predict that about 95\%  of the universal matter is invisible, including dark matter (DM) and dark energy (DE). The unified models of two unknown dark sectors (DM and DE) have been studied in several theories, e.g. in the standard cosmology  \cite{UM1,UM2,UM3}, in the  Ho$\check{r}$ava-Lifshitz  gravity \cite{UM-HL}, in the RS  \cite{UM-RS} and the KK higher-dimension gravity  \cite{UM-KK}. In this paper, we study the unified model of dark components in theory of varying gravitational constant (VG). The attractive point of this model is that  the variation of $G$ could result  to the invisible components in universe, by relating the Lagrangian quantity of the generalized Born-Infeld theory to the VG theory.  One source of DM and DE is introduced.  In addition, given that  cosmic string have been studied in some  fields, such as in emergent universe \cite{cs-emergent,cs-emergent1}, in modified gravity \cite{cs-modified},  in inflation theory \cite{cs-inflation},  and so on \cite{cs-others,cs-others1,cs-others2,cs-others3}. Here we discuss  the effect of  a cosmic string fluid on cosmic parameters  in VG theory. Using the Markov Chain Monte Carlo (MCMC) method \cite{MCMC}, the cosmic constraints on unified model of DM and DE  with (or without) a cosmic string fluid are performed   in the framework of time-varying gravitational constant. The used cosmic data  include the lookback time (LT) data \cite{lt-data,lt-data1},
the model-independent gamma ray bursts (GRBs) data \cite{grbs-data}, the growth function (GF) of matter linear perturbations \cite{fdata1,fdata12,fdata-add1,fdata-add2,fdata2,fdata3,fdata4,fdata5}, the type Ia supernovae (SNIa) data with  systematic  errors \cite{sn-scp-data},  the cosmic microwave background (CMB)  \cite{9ywmap}, and the baryon acoustic oscillation (BAO) data including the radial BAO scale measurement \cite{bao-radial} and the peak-positions  measurement \cite{BAO-WiggleZ,BAO-2dFGRs,BAO-SDSS}.

\section{$\text{A time-varying gravitational constant theory with unified dark sectors and  a cosmic string fluid}$}

We adopt the  Lagrangian quantity of  system \begin{equation}
L=\sqrt{g}\left(\frac{R}{G(t)}+16\pi \mathcal{L}_{u}\right)\label{L-T}
\end{equation}
with a parameterized  time-varying gravitational constant  $G=G_{0}a(t)^{-\beta}$. $t$ is the cosmic time, $a=(1+z)^{-1}$ is the cosmic scale factor, and $z$ denotes the cosmic redshift. $g$ is  the  determinant of metric, $R$ is the Ricci scalar, and $\mathcal{L}_{u}=\mathcal{L}_{b}+\mathcal{L}_{r}+\mathcal{L}_{d}+\mathcal{L}_{s}$ corresponds to the Lagrangian density of universal matter including the visible ingredients: baryon $\mathcal{L}_{b}$ and radiation $\mathcal{L}_{r}$ and the invisible  ingredients: dark sectors $\mathcal{L}_{d}$ and cosmic string (CS) fluid $\mathcal{L}_{s}$. Utilizing the variational principle, the gravitational  field equation can be derived \cite{vg-field},
\begin{equation}
 R_{\mu\nu}-\frac{1}{2}Rg_{\mu\nu}=8\pi GT_{\mu\nu}+G(\nabla_{\mu}\partial_{\nu}G^{-1}-g_{\mu\nu}\nabla_{\sigma}\partial^{\sigma}G^{-1})\label{Einstein-eq}
\end{equation}
in which $R_{\mu\nu}$ is  the Ricci tensor, $T_{\mu\nu}$ is the energy-momentum tensor   of universal   matter that comprise the pressureless baryon ($w_{b}=\frac{p_{b}}{\rho_{b}}=0$), the positive-pressure  photon ($w_{r}=\frac{p_{r}}{\rho_{r}}=\frac{1}{3}$) , the CS fluid ($w_{s}=\frac{p_{s}}{\rho_{s}}=-\frac{1}{3}$)  and the unknown dark components $(w_{d}=\frac{p_{d}}{\rho_{d}})$. $w$ is the equation of state (EoS), $p$ is the pressure  and $\rho$ denotes the energy density, respectively.  Taking the covariant divergence for Eq. (\ref{Einstein-eq}) and utilizing the Bianchi identity result  to
\begin{equation}
 3H(\frac{\dot{G}}{G})^{2}+3\frac{\ddot{a}}{a}\frac{\dot{G}}{G}+8\pi[\dot{G}\rho+G\dot{\rho}+3HG(\rho+p)]=0\label{VG-conservation-eq}
\end{equation}
or its equivalent form
\begin{equation}
 3H\beta[(\beta-1)H^{2}-\dot{H}]+8\pi G[\dot{\rho}+3H(\rho+p)-\beta H \rho]=0.\nonumber
\end{equation}
In the Friedmann-Robertson-Walker geometry, the evolutional equations of universe in VG theory are
\begin{equation}
H^{2}=\frac{8\pi G_{0}}{3}a^{-\beta} \rho-\beta H^{2},\label{H-total}
\end{equation}
\begin{equation}
2\frac{\ddot{a}}{a}+H^{2}=-8\pi G_{0}a^{-\beta}p-\beta H^{2}-\beta^{2} H^{2}-\beta\frac{\ddot{a}}{a}.\label{acc-vg}
\end{equation}
From Eq.(\ref{H-total}), we can see that    a CS fluid can be  equivalent to a curvature term  in constant-$G$ theory, while this fluid could not be equivalent to the curvature term   in the VG theory due to the term $a^{-\beta}$ multiplying the density.   Combing the Eqs. (\ref{VG-conservation-eq}), (\ref{H-total}) and  (\ref{acc-vg}), we have
 \begin{equation}
\dot{\rho}+3H(\rho+\frac{2+2\beta}{2+\beta}p)=\frac{\beta-\beta^{2}}{2+\beta} H\rho.\label{conservation}
\end{equation}
"Dot" represents  the derivative with respect to cosmic time $t$. Integrating Eq. (\ref{conservation}) can gain  the energy density of baryon $\rho_{b}\propto a^{\frac{-\beta^{2}-2\beta-6}{2+\beta}}$, the energy density of radiation
 $\rho_{r} \propto  a^{\frac{-\beta^{2}-4\beta-8}{2+\beta}}$ and the energy density of cosmic string  $\rho_{s} \propto  a^{\frac{-\beta^{2}-4}{2+\beta}}$. Relative to the constant-$G$ theory,   the evolutional equations of energy densities are obviously modified in VG theory for the existence of VG parameter $\beta$.

We concentrate on the Lagrangian density of dark components with the form $\mathcal{L}_{d}=-A^{\frac{1}{1+\alpha}} [1-(V^{'}(\varphi))^{\frac{1+\alpha}{2\alpha}}]^{\frac{\alpha}{1+\alpha}}$ from the  generalized Born-Infeld theory \cite{L-UM}, in which $V(\varphi)$ is the potential.
  Relating this scalar field $\varphi$ with the time-varying gravitational constant by $\varphi(t)=G(t)^{-1}$, it is then found that the dark ingredients can be induced by the variation of $G$.  The energy density of dark  fluid in  $VG$ frame  complies with
\begin{equation}
\rho_{d}=\rho _{0d}[B_s+(1-B_s)a^{(-3+\frac{\beta-\beta^{2}}{2+\beta})(1+\alpha)}]^{\frac 1{1+\alpha }},\label{rho-um}
\end{equation}
here parameter $\beta$ reflects the variation of $G$, $\alpha$ and $B_s=\frac{6+6\beta}{\beta^{2}+2\beta+6}\frac {A}{\rho _{0VG-GCG}^{1+\alpha}}$ are  model parameters.
 Eq. (\ref{rho-um})  shows that the behavior of $\rho_{d}$  is like cold DM at early time\footnote{ $\beta$ describes the effect on energy density of dark matter from variation of $G$.}  (for  $a\ll 1$, $\rho_{d}\approx \rho_{0d}(1-B_{s})^{\frac{1}{1+\alpha}}a^{-3+\frac{\beta-\beta^{2}}{2+\beta}}$),  and like cosmological-constant type DE at late time (for $a\gg 1$, $\rho_{d}\approx \rho_{0d}B_{s}^{\frac{1}{1+\alpha}}$). Then Eq. (\ref{rho-um}) introduces  a unified model (UM) of dark sectors in VG theory (called VG-UM). The Hubble parameter $H$ in the VG-UM model reads
\begin{equation}
H=\sqrt{\frac{H^{2}_{0}}{1+\beta}\{\Omega_{0d}[B_{s}a^{-\beta(1+\alpha)}+(1-B_{s}) a^{-(3+\frac{2\beta^{2}+\beta}{2+\beta})(1+\alpha)}]^{\frac{1}{1+\alpha}}+\Omega_{b}
a^{\frac{-2\beta^{2}-4\beta-6}{2+\beta}}+\Omega_{r}a^{\frac{-2\beta^{2}-6\beta-8}{2+\beta}}+\Omega_{s}a^{\frac{-2\beta^{2}-2\beta-4}{2+\beta}}\}},\label{HE-vg-um}
\end{equation}
with Hubble constant $H_0$ and  dimensionless energy densities $\Omega_{b}= \frac{8 \pi G_{0}\rho_{0b}}{3H^{2}_{0}}$,  $\Omega_{r}= \frac{8\pi G_{0}\rho_{0r}}{3H_{0}^{2}}$,  $\Omega_{s}= \frac{8 \pi G_{0}\rho_{0s}}{3H^{2}_{0}}$, and $\Omega_{0d}+\Omega_{b}+\Omega_{r}+\Omega_{s}=1+\beta$.  For $\beta=0$, bove equations are reduced to the  standard  forms  in the constant-$G$ theory.

\section{$\text{Data fitting}$}

\subsection{$\text{Lookback Time}$}

Refs. \cite{lt-df,lt-df1} define  the LT as the difference between the current age $t_{0}$ of universe  at $z=0$ and the age $t_{z}$ of a light ray  emitted at $z$,
\begin{equation}
t_{L}(z)=\int_{0}^{z}\frac{dz^{'}}{(1+z^{'})H(z^{'})}.\label{eq:tL}
\end{equation}
Then the age $t(z_{i})$ of an object at redshift $z_{i}$ can be expressed by the difference between the age of universe at $z_{i}$ and the age of universe at $z_{F}$ (object was born) \cite{lt-data},
\begin{equation}
t(z_{i})=\int_{z_{i}}^{\infty}\frac{dz^{'}}{(1+z^{'})H(z^{'})}-\int_{z_{F}}^{\infty}\frac{dz^{'}}{(1+z^{'})H(z^{'})}
=t_{L}(z_{F})-t_{L}(z_{i}).\label{eq:ti}
\end{equation}
For an object at redshift $z_{i}$, the observed LT subjects to
\begin{equation}
t_{L}^{obs}=t_{L}(z_{F})-t_{L}(z_{i})=[t_{0}^{obs}-t(z_{i})]-[t_{0}^{obs}-t_{L}(z_{F})]=t_{0}^{obs}-t(z_{i})-df.
\label{eq:t-obs}
\end{equation}
One defines
\begin{equation}
 \chi^{2}_{age}=\sum_{i}\frac{[t_{L}(z_{i})-t_{L}^{obs}(z_{i},df)]^{2}}{\sigma_{T}^{2}}+\frac{[t_{0}-t_{0}^{obs}]^{2}}{\sigma^{2}_{t_{0}^{obs}}},
\label{eq:chi-age}
\end{equation}
with $\sigma_{t_{0}^{obs}}^{2}+\sigma_{i}^{2}=\sigma_{T}^{2}$. $\sigma_{t_{0}^{obs}}$ is the uncertainty of the total universal  age, and $\sigma_{i}$ is the uncertainty of the LT of galaxy  $i$. Marginalizing the 'nuisance' parameter $df$ results to  \cite{lt-chi2}
\begin{equation}
 \chi^{2}_{LT}(p_{s})=-2\ln\int_{0}^{\infty}d(df)\exp(-\chi^{2}_{age}/2)=A-\frac{B^{2}}{C}+\frac{[t_{0}-t_{0}^{obs}]^{2}}{\sigma^{2}_{t_{0}^{obs}}}-2\ln[\sqrt{\frac{\pi}{2C}} erfc(\frac{B}{\sqrt{2C}})],
\label{eq:chi-LT}
\end{equation}
where
$A=\sum_{i} \frac{\Delta^{2}}{\sigma_{T}^{2}}, B=\sum_{i} \frac{\Delta}{\sigma_{T}^{2}},C=\sum_{i} \frac{1}{\sigma_{T}^{2}}$ and  $\Delta=t_{L}(z_{i})-[t_{0}^{obs}-t(z_{i})]$, respectively. $p_{s}$ denotes the  theoretical  model parameters. erfc($x$) = 1-erf($x$) is the complementary error function of $x$. The observational universal age at today $t_{0}^{obs}=13.75\pm 0.13$  Gyr \cite{lt-age} is used, and the observational data  on the  galaxies age are listed in table \ref{LT-data}.

\begin{table}[!htbp]
 \vspace*{-12pt}
 \begin{center}
 \begin{tabular}{  | c||  c |  c |  c |  c |  c  |  c |  c|  c|  c|  c|  c|  c |  c|  c|  c|  c|  c|  c|  c |} \hline\hline
$z_{i}$ &0.10 &0.25&0.60 &0.70 &0.80 &1.27&0.1171&0.1174&0.222 &0.2311&0.3559&0.452 &0.575 &0.644 &0.676 &0.833 &0.836 &0.922 &1.179\\\hline
$t_{i}$ &10.65 &8.89&4.53 &3.93 &3.41 &1.60&10.2  & 10.0 &  9.0 &  9.0 & 7.6  &  6.8 &7.0   & 6.0  &6.0   &6.0   &5.8   &5.5   &4.6  \\\hline\hline
$z_{i}$ &1.222&1.224&1.225&1.226&1.34&1.38&1.383 &1.396 &1.43  &1.45  &1.488 &1.49  &1.493 &1.51  &1.55  &1.576 &1.642 &1.725 &1.845 \\\hline
$t_{i}$ &3.5  &4.3  &3.5  &3.5  &3.4  &3.5&3.5   &3.6   &3.2   &3.2   &3.0   &3.6   &3.2   &2.8   &3.0   &2.5   &3.0   &2.6   &2.5   \\\hline\hline
 \end{tabular}
 \end{center}
 \caption{The 38 data points of galaxy age \cite{lt-data,lt-data1}. The first 6 data are from Ref.\cite{lt-data}. }\label{LT-data}
 \end{table}

\subsection{$\text{Gamma Ray Bursts}$}
 In GRBs observation, the famous Amati's correlation is  $\log\frac{E_{iso}}{erg}=a+b \log\frac{E_{p,i}}{300keV}$ \cite{grbs-31,grbs-33}, where $E_{iso}=4\pi d_{L}^{2}S_{solo}/(1+z)$  and $E_{p,i} = E_{p,obs}(1+z)$ are the isotropic energy and the cosmological rest-frame spectral peak energy, respectively. $d_{L}$ is the luminosity distance and  $S_{bolo}$ is the bolometric fluence of GRBs. Ref. \cite{grbs-36} introduced a  model-independent quantity of distance measurement,
\begin{equation}
\overline{r}_{p}(z_{i})= \frac{r_{p}(z)}{r_{p}(z_{0})},r_{p}(z)= \frac{(1+z)^{1/2}}{z}\frac{H_{0}}{c}r(z), r(z)=\frac{d_{L}(z)}{1+z} \label{eq:grbs-3}
\end{equation}
with $z_{0}$ being the lowest GRBs redshift. For GRBs constraint, $\chi^{2}_{GRBs}$ has a form
\begin{eqnarray}
\chi^{2}_{GRBs}(p_{s})=[\Delta \overline{r}_{p}(z_{i})]\cdot(Cov^{-1}_{GRBs})_{ij}\cdot [\Delta \overline{r}_{p}(z_{i})] \label{eq:grbs-chi}
\end{eqnarray}
in which $\Delta \overline{r}_{p}(z_{i})=\overline{r}^{data}_{p}(z_{i})-\overline{r}_{p}(z_{i})$,  and ($Cov^{-1}_{GRBs})_{ij}$ is the covariance matrix.  Using 109 GRBs data, Ref. \cite{grbs-data} obtained $5$ model-independent datapoints  listed in table \ref{table-grb-data}, where $\sigma(\overline{r}_{p}(z_{i}))^{+}$ and   $\sigma(\overline{r}_{p}(z_{i}))^{-}$ are the $1\sigma$ errors. The $\{\overline{r}_{p}(z_{i})\}$ correlation matrix is  \cite{grbs-data}
\begin{eqnarray}
&&(\overline{Cov}_{GRB})= \left(\begin{array}{c}
 1.0000~~~0.7780~~~0.8095~~~0.6777~~~0.4661\\
 0.7780~~~1.0000~~~0.7260~~~0.6712~~~0.3880\\
 0.8095~~~0.7260~~~1.0000~~~0.6046~~~0.5032\\
 0.6777~~~0.6712~~~0.6046~~~1.0000~~~0.1557\\
 0.4661~~~0.3880~~~0.5032~~~0.1557~~~1.0000
\end{array}\right),\label{cov-GRB}
\end{eqnarray}
with the covariance matrix
\begin{equation}
(Cov_{GRB})_{ij}=\sigma(\overline{r}_{p}(z_{i}))\sigma(\overline{r}_{p}(z_{j}))(\overline{Cov}_{GRB})_{ij},\label{eq:cov-grb}
\end{equation}
where
$\sigma(\overline{r}_{p}(z_{i}))=\sigma(\overline{r}_{p}(z_{i}))^{+}$, if $\overline{r}_{p}(z)\geq \overline{r}_{p}(z)^{data}$;
$\sigma(\overline{r}_{p}(z_{i}))=\sigma(\overline{r}_{p}(z_{i}))^{-}$, if $\overline{r}_{p}(z)< \overline{r}_{p}(z)^{data}$.

\begin{table}[ht]
\begin{center}
\begin{tabular}{|c||l|l|l|l|}
\hline\hline  Number   & $z$   & $\overline{r}_{p}^{data}(z)$ &  $\sigma(\overline{r}_{p}(z_{i}))^{+}$    &  $\sigma(\overline{r}_{p}(z_{i}))^{-}$
\\ \hline 0        & 0.0331  &1.0000 &---- &----
\\ \hline 1        &1.0000   &0.9320 &0.1711 & 0.1720
\\ \hline 2        & 2.0700  &0.9180 &0.1720 & 0.1718
\\ \hline 3        &3.0000   &0.7795 &0.1630 & 0.1629
\\ \hline 4        &4.0480   &0.7652 &0.1936 &0.1939
\\ \hline 5        &8.1000   &1.1475 &0.4297 &0.4389
\\ \hline \hline
\end{tabular}
\end{center}
\caption{\label{table-grb-data}
 Distances calculated by using the 109 GRBs data via Amati's correlation \cite{grbs-data}.}
\end{table}

\subsection{$\text{Growth Function of Matter Linear Perturbations}$}

The $ \chi^2_{GF}$ can be constructed by the growth function  of matter linear perturbations $f$
 \begin{equation}
 \chi_{GF}^2(p_{s})=\sum_{i} \frac{[f_{th}(p_{s},z_i)-f_{obs}(z_i)]^2}{\sigma^2(z_i)},\label{chi2-gf}
 \end{equation}
where the used observational values of $f_{obs}$ are listed in table \ref{table-f-data}. $f$ is defined via $f(a)=\frac{aD^{'}(a)}{D(a)}$, with $D=\frac{\frac{\delta \rho}{\rho}(a)}{\frac{\delta \rho}{\rho}(a=1)}$. $'$ denotes derivative with respect to $a$. So in theory, $f$ can be gained  by  solving the following differential equation in VG theory
\begin{equation}
 D^{''}(a)+[\frac{H^{'}(a)}{H(a)}+\frac{1}{a}+\frac{4+2\beta+2\beta^{2}}{a(2+\beta)}]D^{'}(a)- \frac{6+2\beta+\beta^{2}}{(2+\beta)^{2}}\frac{H_{0}^{2}\Omega_{0m}}{H(a)^{2}a^{2}}a^{\frac{-6-2\beta-\beta^{2}}{2+\beta}}D(a)=0.\label{D-evolutoin-text}
 \end{equation}
For $\beta=0$, above equation reduces to the constant-$G$ theory. The derivation of evolutional equation $D(a)$ in VG theory are shown in appendix. Comparing  with the most popular $\Lambda$CDM model,  the effective current matter density can be written, $\Omega_{0m}=\Omega_{b}+(1+\beta-\Omega_{s}-\Omega_{b}-\Omega_{r})(1-B_{s})$ for VG-UM.   Obviously,  for $\beta=0$ it is consistent with the  form of $\Omega_{0m}$ in UM of constant-$G$ theory \cite{om-gcg-eff0,om-gcg-eff01,om-gcg-eff02}.

\begin{table}[ht]
\begin{center}
\begin{tabular}{|c||l|l|l|l|l|l|l|l|l|l|} \hline\hline
$z_{i}$           &0.15 & 0.22 & 0.32   &0.35  & 0.41  &0.55   & 0.60    & 0.77  &0.78 & 1.4\\ \hline
$f_{obs}$  &$0.51 \pm 0.11$  &$0.60  \pm 0.10$ &$0.654 \pm 0.18$ &$0.70  \pm 0.18$ &$0.50  \pm 0.07$ &$0.75  \pm 0.18$
                  &$0.73  \pm 0.07$ & $0.91  \pm 0.36$  &$0.70  \pm  0.08$ &$0.90  \pm 0.24$ \\ \hline
 Ref.       &\cite{fdata1,fdata12} &\cite{{fdata-add1}} &\cite{{fdata-add2}} &\cite{fdata2} &\cite{{fdata-add1}}
                 &\cite{fdata3} &\cite{{fdata-add1}} &\cite{fdata4} &\cite{{fdata-add1}} &\cite{fdata5}\\ \hline \hline
\end{tabular}
\end{center}
\caption{The observational data of growth function $f_{obs}$.\label{table-f-data} }
\end{table}

\subsection{$\text{Type Ia Supernovae}$}
We use the Union2 dataset of SNIa  published in Ref.  \cite{sn-scp-data}. In VG theory, the theoretical distance modulus $\mu_{th}(z)$ is written as $\mu_{th}(z)=5\log_{10}[D_{L}(z)]+\frac{15}{4}\log_{10}\frac{G}{G_{0}}+\mu_{0}$, where $D_{L}(z)=\frac{H_{0}}{c}(1+z)^{2}D_A(z)$  and  $\mu_{0}=5log_{10}(\frac{H_{0}^{-1}}{Mpc})+25=42.38-5log_{10}h$. $h$ is a re-normalized quantity defined by  $H_0 =100 h~{\rm km ~s}^{-1} {\rm Mpc}^{-1}$. $D_A(z)=\frac{c}{(1+z)\sqrt{|\Omega_k|}}\mathrm{sinn}[\sqrt{|\Omega_k|}\int_0^z\frac{dz'}{H(z')}]$ is the proper  angular diameter distance, here  $\mathrm{sinn}(\sqrt{|\Omega_k|}x)$ denotes $\sin(\sqrt{|\Omega_k|}x)$, $\sqrt{|\Omega_k|}x$ and $\sinh(\sqrt{|\Omega_k|}x)$ for $\Omega_k<0$, $\Omega_k=0$ and
$\Omega_k>0$, respectively. Cosmic constraint  from SNIa observation can be done by a  calculation  on \cite{chi2-SNIa,chi2-SNIa1,chi2-SNIa2,chi2-SNIa3,chi2-SNIa4,chi2-SNIa5,chi2-SNIa6,chi2-SNIa7,chi2-SNIa8,chi2-SNIa9}
\begin{eqnarray}
\chi^2_{SNIa}(p_{s})  =  \sum_{SNIa}\frac{\left\{\mu_{th}(p_{s},z_i)-\mu_{obs}(z_i)\right\}^2} {\sigma_{\mu_{i}}^2}
=\sum_{SNIa}\frac{\{5\log_{10}[D_{L}(p_{s},z)]+\frac{15}{4}\log_{10}\frac{G}{G_{0}}-m_{obs}(z_{i})+M^{'}\}^{2}}{\sigma_{i}^{2}},\label{eq:chi2SN1}
\end{eqnarray}
where $\mu_{obs}(z_{i})=m_{obs}(z_{i})-M$ is the observed distance moduli, with the absolute magnitude $M$.  The nuisance parameter $M^{'}=\mu_{0}+M$  can be marginalized over analytically, $\bar{\chi}_{SNIa}^2(p_s) = -2 \ln \int_{-\infty}^{+\infty}\exp \left[-\frac{1}{2} \chi_{SNIa}^2(p_s,M^{\prime}) \right] dM^{\prime}$
resulting to  \cite{sn-chi2abc,sn-chi2abc1,sn-chi2abc2,sn-chi2abc3,sn-chi2abc4,sn-chi2abc5,sn-chi2abc6,sn-chi2abc7,Xu1,Xu2}
\begin{equation}
\chi^2_{SNIa}(p_s)=A-(B^2/C),\label{eq:chi2SN}
\end{equation}
where
\begin{eqnarray}
&&A=\sum_{SNIa}\{5\log_{10}[D_{L}(p_{s},z_{i})]+\frac{15}{4}\log_{10}\frac{G}{G_{0}}-m_{obs}(z_{j})\} \cdot C_{ij}^{-1}\cdot  \{5\log_{10}[D_{L}(p_{s},z_{j})]+\frac{15}{4}\log_{10}\frac{G}{G_{0}}-m_{obs}(z_{j})\}\nonumber\\
&&B=\sum_{SNIa}C_{ij}^{-1}\cdot \{5\log_{10}[D_{L}(p_{s},z_{j})]+\frac{15}{4}\log_{10}\frac{G}{G_{0}}-m_{obs}(z_{j})\}\nonumber\\
&&C=\sum_{SNIa}C_{ii}^{-1}.\label{eq:snABC-sys}
\end{eqnarray}
The inverse of covariance matrix  $C_{ij}^{-1}$ with systematic errors can be found in Refs. \cite{sn-scp-data,sn-web}.

\subsection{$\text{Cosmic Microwave Background}$}

 $\chi^{2}_{CMB}$ has a form \cite{Xu3,Wu3}
\begin{eqnarray}
&&\chi^2_{CMB}(p_{s})=\bigtriangleup d_i[Cov^{-1}(d_i(p_{s}),d_j(p_{s}))][\bigtriangleup
d_i]^t,\label{chi2-CMB}
\end{eqnarray}
with $\bigtriangleup d_{i}(p_{s})=d_{i}^{theory}(p_{s})-d_{i}^{obs}$. 9-year WMAP gives $d_{i}^{obs}=[l_{A}(z_{\ast})=302.04, R(z_{\ast})=1.7246, z_{\ast}=1090.88]$,
and the corresponding inverse covariance matrix \cite{9ywmap}
\begin{eqnarray}
&&Cov^{-1}= \left(\begin{array}{c}
 3.182~~~18.253~~~-1.429\\
 18.253~~~11887.879~~~-193.808\\
 -1.429~~~-193.808~~~4.556
\end{array}\right).\label{cov-CMB}
\end{eqnarray}
 $z_{\ast}=1048\left[1+0.00124(\Omega_bh^2)^{-0.738}\right]\left[1+g_1(\Omega_{0m} h^2)^{g_2}\right]$ is the redshift at decoupling epoch of photons
with $g_1=0.0783(\Omega_bh^2)^{-0.238}\left(1+ 39.5(\Omega_bh^2)^{0.763}\right)^{-1}$ and $g_2=0.560\left(1+ 21.1(\Omega_bh^2)^{1.81}\right)^{-1}$, $l_A(p_{s};z_{\ast})=(1+z_{\ast})\frac{\pi D_A(p_{s};z_{\ast})}{r_s(z_{\ast})}$ is the acoustic scale, and  $R(p_{s};z_{\ast})=\sqrt{\Omega_{0m} H^2_0}(1+z_{\ast})D_A(p_{s};z_{\ast})/c$ is the CMB shift parameter.

\subsection{$\text{Baryon Acoustic Oscillation}$}
The radial (line-of-sight) BAO scale measurement from galaxy power spectra can be depicted by
\begin{equation}
\Delta z_{BAO}(z)=\frac{H(z)r_{s}(z_{d})}{c}.
\end{equation}
Two  observational  values are $\Delta z_{BAO}(z=0.24)=0.0407\pm 0.0011$ and $\Delta z_{BAO}(z=0.43)=0.0442\pm 0.0015$, respectively \cite{bao-radial}.
Here $r_s(z)$ is the comoving sound horizon size $r_s=c\int_0^t\frac{c_sdt}{a}$. $c_s$ is the sound speed of the photon$-$baryon fluid, $c_s^{-2}=3+\frac{4}{3}\times(\frac{\Omega_{b}}{\Omega_{\gamma})})a$.  $z_d$ denotes the drag epoch, $z_d=\frac{1291(\Omega_{0m}h^2)^{-0.419}}{1+0.659(\Omega_{0m}h^2)^{0.828}}[1+b_1(\Omega_{b}h^2)^{b_2}]$ with
$b_1=0.313(\Omega_{0m}h^2)^{-0.419}[1+0.607(\Omega_{0m}h^2)^{0.674}]$ and $b_2=0.238(\Omega_{0m}h^2)^{0.223}$.

The measurement of BAO peak positions can be performed by the WiggleZ Dark Energy Survey \cite{BAO-WiggleZ}, the Two Degree Field Galaxy Redshift Survey \cite{BAO-2dFGRs} and the Sloan Digitial Sky Survey \cite{BAO-SDSS}. Introducing $ D_V(z)=[(1+z)^2 D^{2}_{A}(z) \frac{cz}{H(z;p_{s})}]^{1/3}$, one can exhibit the observational data from BAO peak positions
\begin{eqnarray}
X= \left(
\begin{array}{c}
 \frac{r_s(z_d)}{D_V(0.106)}-0.336 \\
 \frac{r_s(z_d)}{D_V(0.2)}-0.1905 \\
 \frac{r_s(z_d)}{D_V(0.35)}-0.1097 \\
  \frac{r_s(z_d)}{D_V(0.44)}-0.0916 \\
   \frac{r_s(z_d)}{D_V(0.6)}-0.0726 \\
    \frac{r_s(z_d)}{D_V(0.73)}-0.0592
\end{array}
\right),
&&V^{-1}= \left(
\begin{array}{cccccc}
 4444 & 0 & 0 & 0 & 0 & 0 \\
 0 & 30318 & -17312 & 0 & 0 & 0 \\
 0 & -17312 & 87046 & 0 & 0 & 0 \\
 0 & 0 & 0 & 23857  & -22747 & 10586 \\
 0 & 0 & 0 & -22747 & 128729 & -59907 \\
 0 & 0 & 0 & 10586 & -59907 & 125536
\end{array}
\right)
\end{eqnarray}
where $V^{-1}$ is  the inverse covariance matrix shown in Ref. \cite{BAO-comatrix}.

The $\chi^{2}_{BAO}$ can be constructed
\begin{equation}
\chi^{2}_{BAO}(p_{s})=\frac{[\Delta z_{BAO}(z=0.24)-0.0407]^{2}}{0.0011^{2}}+\frac{[\Delta z_{BAO}(z=0.43)-0.0442]^{2}}{0.0015^{2}}+X^tV^{-1}X.
\end{equation}
  $X^t$ denotes the  transpose of $X$.

\section{$\text{Cosmic constraints on unified model of dark sectors with  (or without) a CS fluid  in  VG theory}$}

Multiplying the separate likelihoods $L_{i}\propto e^{-\chi^{2}_{i}/2}$, one can express the joint analysis of $\chi^{2}$
\begin{equation}
\chi^{2}=\chi^{2}_{LT}+\chi^{2}_{GRBs}+\chi^{2}_{GF}+\chi^{2}_{SNIa}+\chi^{2}_{CMB}+\chi^{2}_{BAO}.\label{chi-total}
\end{equation}

\subsection{$\text{The case with a CS fluid }$}

\begin{figure}[ht]
  \includegraphics[width=12cm]{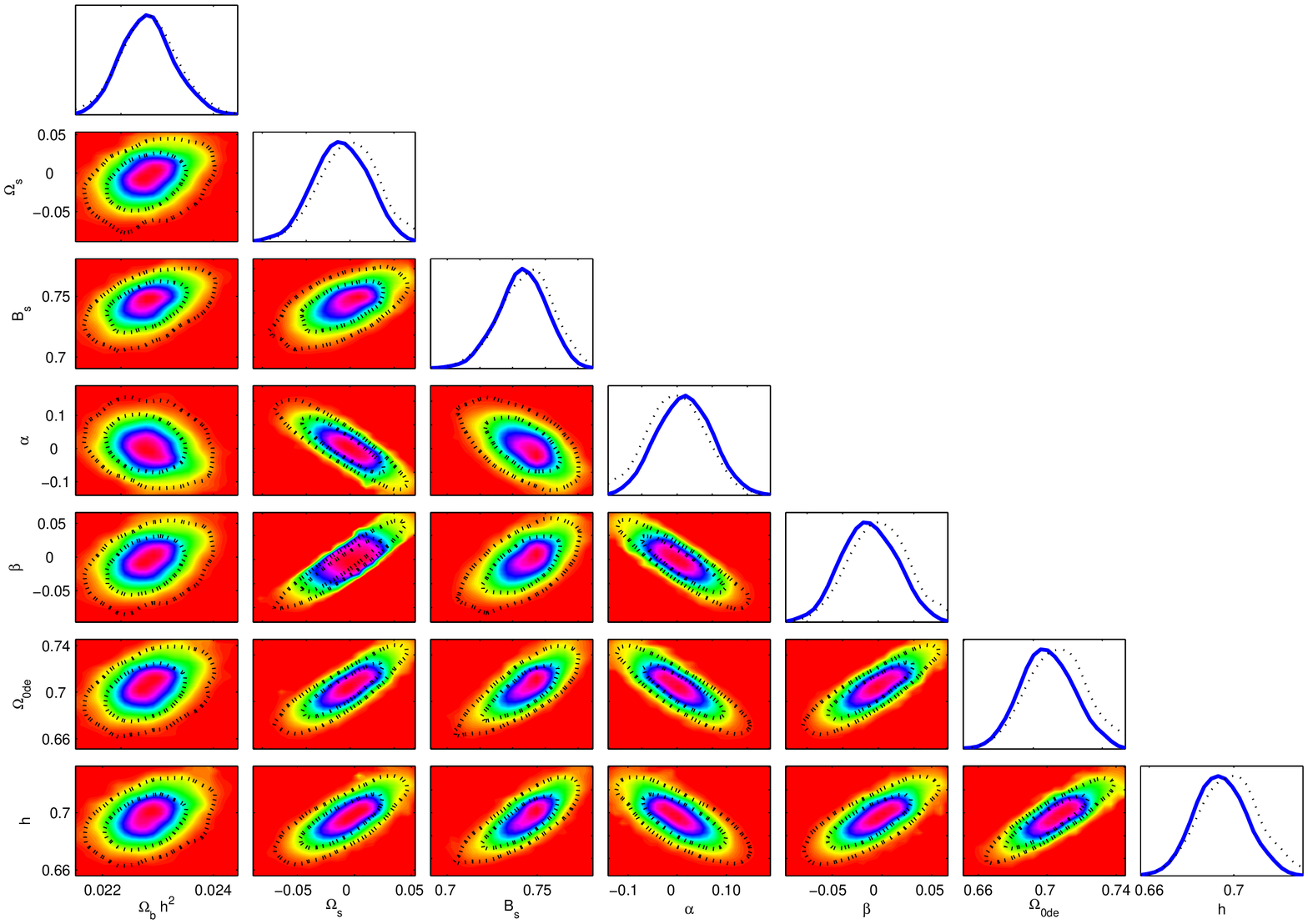}
  \includegraphics[width=5cm]{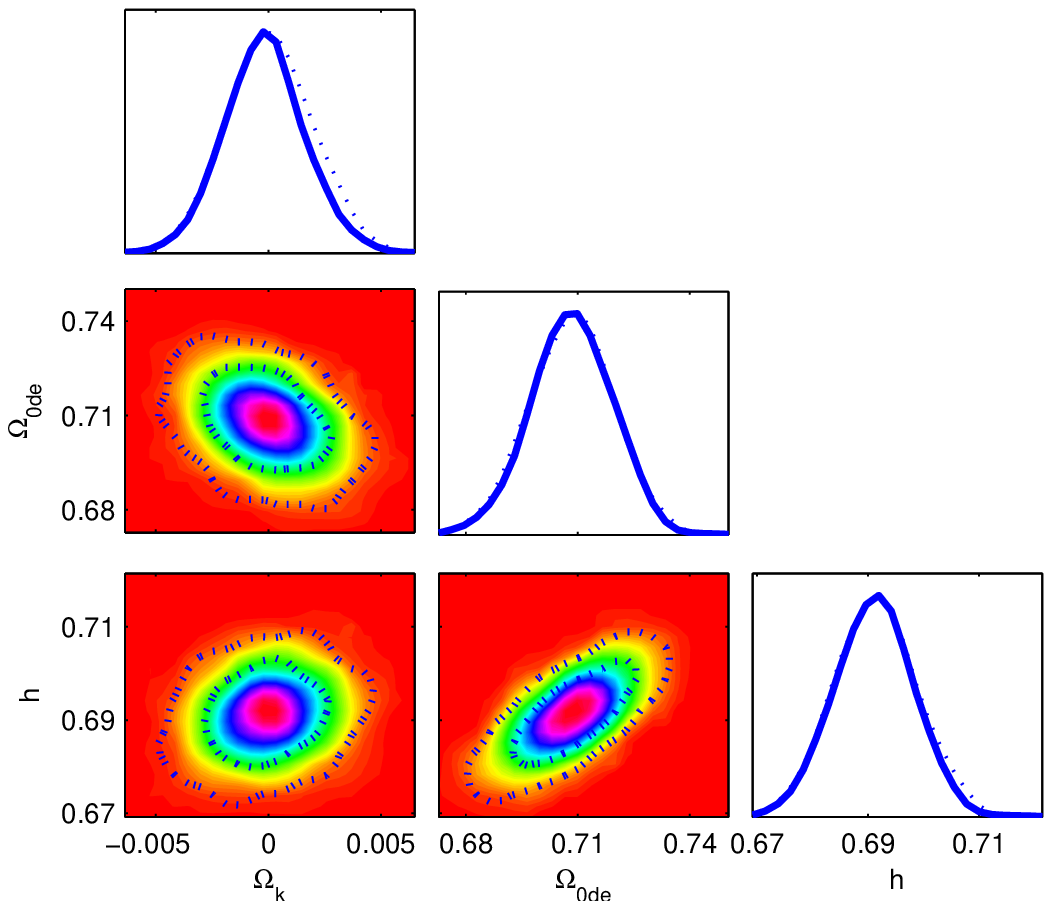}\\
  \caption{ $1\sigma$ and $2\sigma$ contours of  parameters for the VG-UM model  with a CS fluid (left) and the  $\Lambda$CDM (right) model.}\label{figure-ab-vgum0-with}
\end{figure}

\begin{table}[!htbp]
 \vspace*{-12pt}
 \begin{center}
 \begin{tabular}{|c|| c| c| c| c|  } \hline\hline
 & Mean values with limits (VG-UM)   & Best fit (VG-UM)  & Mean values with limits ($\Lambda$CDM) &  Best fit ($\Lambda$CDM) \\\hline
----  &  $\Omega_{s}=-0.0106^{+0.0312+0.0582}_{-0.0305-0.0509}$            & 0.0006
                  & $\Omega_{k}=-0.0002^{+0.0024+0.0052}_{-0.0024-0.0048}$             &  -0.0004   \\\hline
 $\beta$          &$-0.0128^{+0.0394+0.0756}_{-0.0385-0.0718}$              &0.0005
                   &  0  & 0 \\\hline 
  $B_{s}$         &$0.7457^{+0.0147+0.0269}_{-0.0145-0.0299}$               &0.7520
                  &  ---- & ---- \\\hline                 
  $\alpha$        &$0.0216^{+0.0757+0.1504}_{-0.0781-0.1466}$       &0.0004
                   & 0  & 0 \\\hline  
   $h$             &$0.6922^{+0.0149+0.0306}_{-0.0149-0.0289}$      &0.6981
                    & $0.6916^{+0.0100+0.0197}_{-0.0101-0.0193}$    & 0.6930
                     \\\hline  
  $100\Omega_{b} h^{2}$
                        &$2.2580^{+0.0555+0.1154}_{-0.0557-0.1051}$      & 2.2691
                        & $2.2683^{+0.0412+0.0815}_{-0.0420-0.0776}$  & 2.266 \\\hline  
  $\Omega_{0de}$        &$0.6983^{+0.0165+0.0347}_{-0.0161-0.0309}$  &0.7175
                        & $0.7098^{+0.0144+0.0265}_{-0.0140-0.0294}$  & 0.7126 \\\hline\hline
 \end{tabular}
 \end{center}
 \caption{The mean values  with  limits and the best-fit values of parameters for VG-UM  model with a CS fluid. }\label{table-vgum0-with}
 \end{table}

In order to obtain the stringent  constraint on VG theory, we utilize  the cosmic data  different from Ref. \cite{vg-field} to calculate the joint likelihood.
 Concretely, the LT data,  the GRBs data, the GF data,  the SNIa data with systematic error and the BAO data from  radial measurement are not used in  Ref. \cite{vg-field}.  After calculation, the 1-dimension distribution and the 2-dimension  contours  of parameters for the  VG-UM model with a CS fluid  are illustrated in  Fig. \ref{figure-ab-vgum0-with}.  From Fig. \ref{figure-ab-vgum0-with} and table \ref{table-vgum0-with}, we can see that the  restriction on dimensionless energy density of CS  is  $\Omega_{s}=-0.0106^{+0.0312+0.0582}_{-0.0305-0.0509}$ in the varying-$G$ theory with containing unified dark sectors. In the constant-$G$ theory,  one knows that a CS fluid  with $w_{s}=-1/3$ is usually equivalent to a  curvature term. But, in the VG theory this equivalence is lost due to the term $a^{-\beta}$ multiplying the density, as shown in Eq.(\ref{H-total}). Comparing the VG theory with the constant-$G$ theory,   it can be seen that the uncertainty of $\Omega_{s}$ in VG theory is larger than some results on  $\Omega_{k}$ in constant-$G$ theory. For example,  using the same data to constrain other models  we have $\Omega_{k}=-0.0002^{+0.0024+0.0052}_{-0.0024-0.0048}$  (with model parameter $\Omega_{0de}=0.7098^{+0.0144+0.0265}_{-0.0140-0.0294}$)  in $\Lambda$CDM model, $\Omega_{k}=-0.0001^{+0.0025+0.0052}_{-0.0025-0.0050}$  (with  model parameters  $B_{s}=0.7665^{+0.0101+0.0194}_{-0.0099-0.0205}$ and  $\alpha=0.0209^{+0.0186+0.0401}_{-0.0189-0.0373}$) in constant-$G$ UM. Taking the $\Lambda$CDM model as a reference, we can see that the influence on the fitting value of  $\Omega_{k}$ is small from the added parameter $B_{s}$ and $\alpha$ as seen in constant-$G$ UM model, while the  influence on the value of $\Omega_{s}$ is large by the added VG parameter $\beta$ as indicated in VG-UM model.  From table  \ref{table-vgum0-with}, one reads VG parameter $\beta=-0.0128^{+0.0394+0.0756}_{-0.0385-0.0718}$. Other parameters   are $B_{s}=0.7457^{+0.0147+0.0269}_{-0.0145-0.0299}$ and $\alpha=0.0216^{+0.0757+0.1504}_{-0.0781-0.1466}$. We then find  at $1\sigma$ confidence level, the flat $\Lambda$CDM model  ($\Omega_{s}=0$, $\beta=0$ and $\alpha=0$) is included in  the VG-UM model  with a CS fluid. This result in VG theory is same as the popular point that the complicated cosmological model is usually degenerate with the $\Lambda$CDM model.

\subsection{$\text{The case without a CS fluid}$}

\begin{figure}[ht]
  \includegraphics[width=12cm]{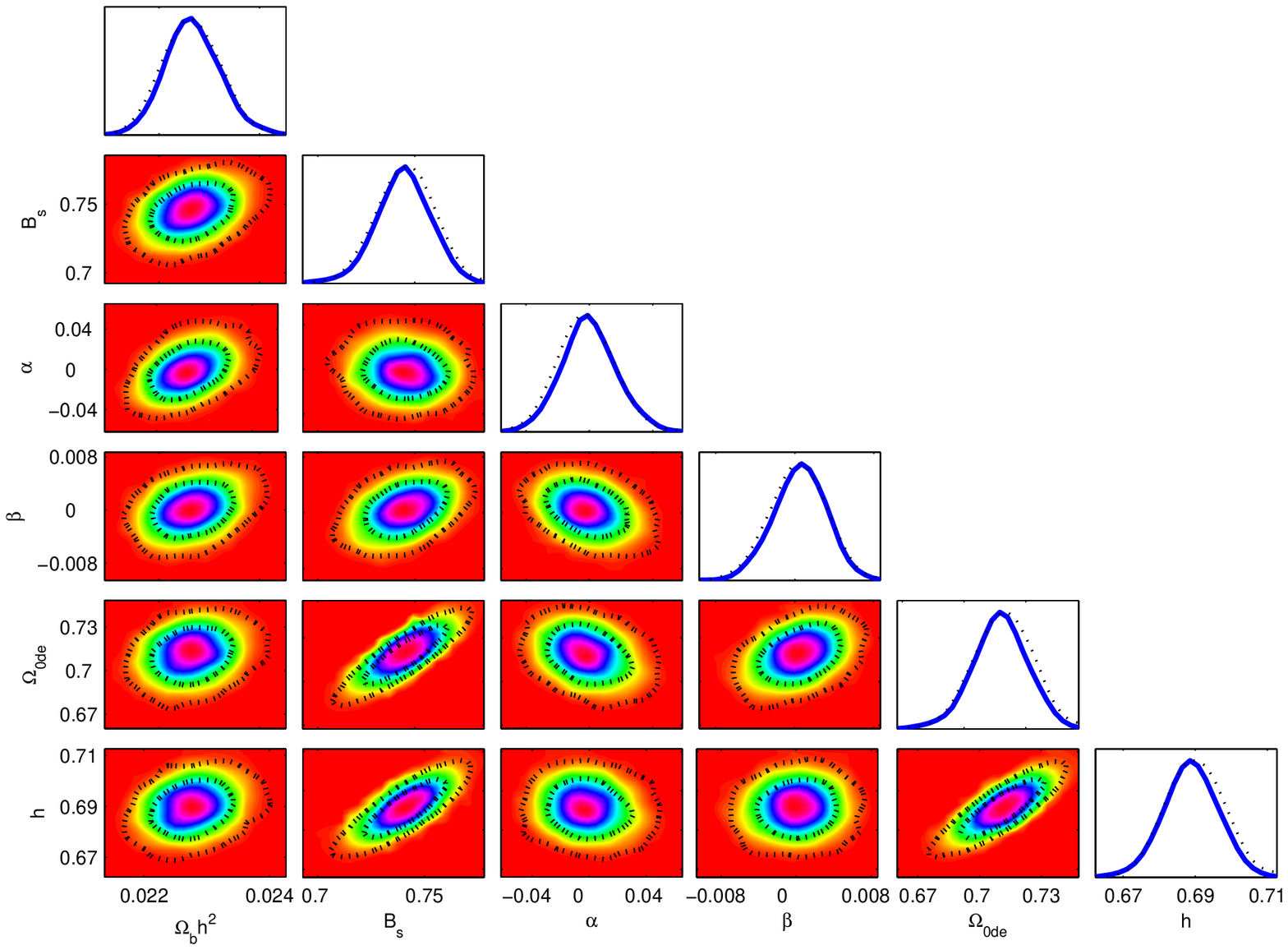}
  \includegraphics[width=5cm]{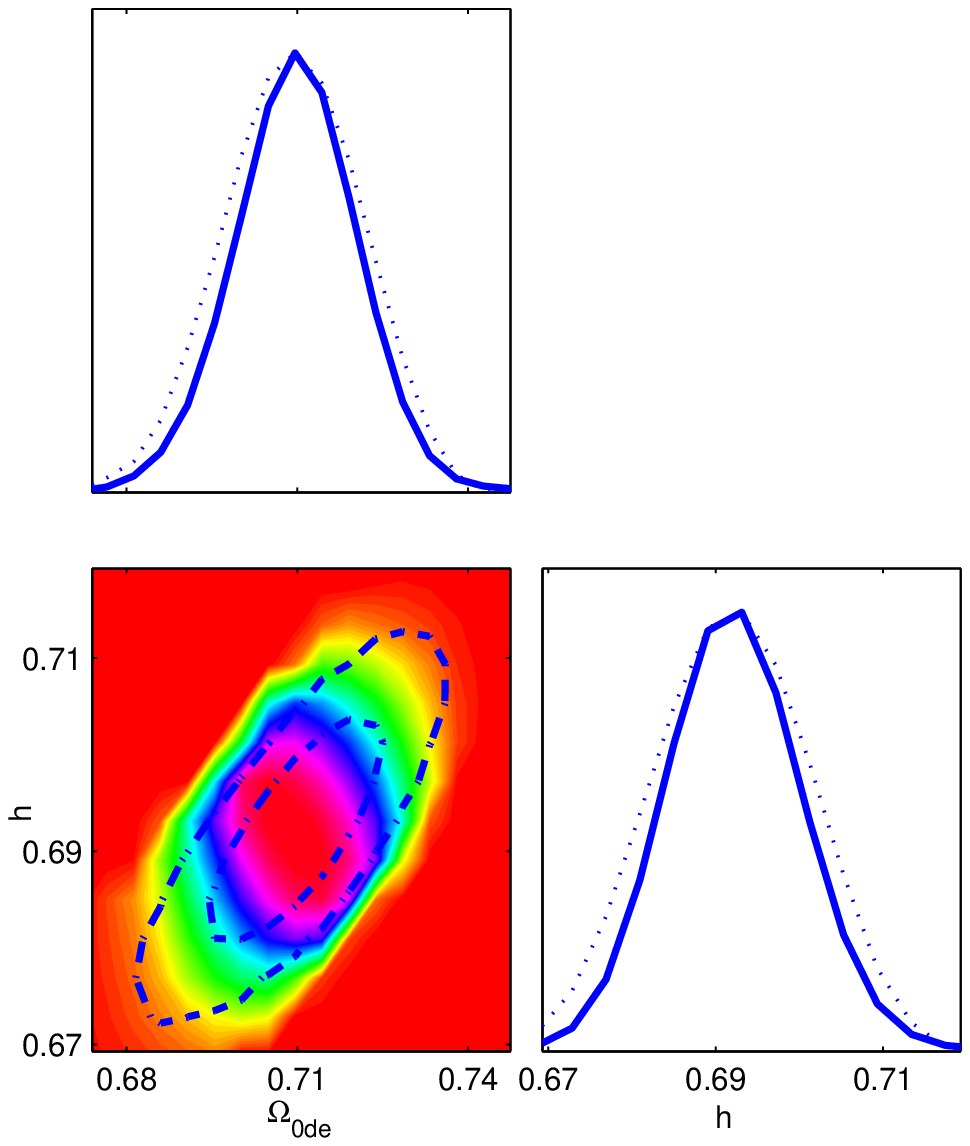}\\
  \caption{ $1\sigma$ and $2\sigma$ contours of parameters for the VG-UM model without a CS fluid (left) and the $\Lambda$CDM (right) model.}\label{figure-ab-vgum0-without}
\end{figure}

\begin{table}[!htbp]
 \vspace*{-12pt}
 \begin{center}
 \begin{tabular}{|c|| c| c| c| c|  } \hline\hline
 & Mean values with limits (VG-UM)   & Best fit (VG-UM)  & Mean values with limits ($\Lambda$CDM) &  Best fit ($\Lambda$CDM) \\\hline
 $\beta$
                  &$0.0007^{+0.0032+0.0062}_{-0.0033-0.0067}$ 
                  &0.0010 &  0  & 0 \\\hline 
  $B_{s}$         &$0.7442^{+0.0137+0.0262}_{-0.0132-0.0292}$ 
                   &0.7440  &  ---- & ---- \\\hline                 
  $\alpha$        &$0.0002^{+0.0206+0.0441}_{-0.0209-0.0422}$ 
                      &0.0073  & 0  & 0 \\\hline  
   $h$             &$0.6905^{+0.0098+0.0191}_{-0.0096-0.0203}$ 
                   &0.6902  & $0.6925^{+0.0094+0.0207}_{-0.0104-0.0198}$    & 0.6923
                     \\\hline  
  $100\Omega_{b} h^{2}$
                        &$2.267^{+0.054+0.116}_{-0.051-0.102}$ 
                        & 2.256 & $2.2647^{+0.0398+0.0789}_{-0.0394-0.0781}$  & 2.262 \\\hline  
  $\Omega_{0de}$
                        &$0.7093^{+0.0148+0.0296}_{-0.0150-0.0309}$ 
                          &0.7095   & $0.7101^{+0.0126+0.0270}_{-0.0135-0.0282}$  & 0.7106 \\\hline\hline
 \end{tabular}
 \end{center}
 \caption{The mean values  with  limits and the best-fit values of model parameters for VG-UM model without a CS fluid. }\label{table-vgum0-without}
 \end{table}

  For the case without a CS fluid,   a stringent constraint  on VG parameter is $\beta=0.0007^{+0.0032+0.0062}_{-0.0033-0.0067}$, where
a small uncertainty   at $2\sigma$ regions for $\beta$ is given. Still, it is shown that the value of  $\beta$ is around zero at $1\sigma$ confidence level for both cases: including or not including a CS  fluid, and the case containing a CS fluid has a larger error for $\beta$   than that not containing a CS fluid.  In VG theory, the constraint  on UM model parameters are
$B_{s}=0.7442^{+0.0137+0.0262}_{-0.0132-0.0292}$, $\alpha=0.0002^{+0.0206+0.0441}_{-0.0209-0.0422}$, $h=0.6905^{+0.0098+0.0191}_{-0.0096-0.0203}$ and $100\Omega_bh^{2}=2.267^{+0.054+0.116}_{-0.051-0.102}$. At $1\sigma$ confidence level, the value of $\alpha=0$ is not excluded, which demonstrates that the $\Lambda$CDM model can not be distinguished from VG-UM model by the joint  cosmic data. Besides the mean values  with limits, the best-fit values of VG-UM model parameters are determined  and exhibited in  table \ref{table-vgum0-without}, too.  As a reference, the $\Lambda$CDM model is  calculated by using  the combined observational data  appeared  in section III, and the best-fit values  and the mean values with limits on $\Lambda$CDM model  are laid in  table \ref{table-vgum0-without}. In $\Lambda$CDM model, one receives $\Omega_{0de}=0.7101^{+0.0126+0.0270}_{-0.0135-0.0282}$  that is  compatible to the effective result of $\Omega_{0de}$ in VG-UM model.

In order to agglomerate and form structure of universe,  one knows that the  baryonic  (and DM) component must have a near zero pressure. Given that $w_{b}=\frac{p_{b}}{\rho_{b}}=\frac{-\beta(1-\beta)}{3(2+\beta)} \sim 0$, $\beta\sim 0$ or $\beta \sim 1$ could be solved. From above constraint on parameter $\beta$, one can see that the solution $\beta\sim 0$ is consistent with our fitting result for both cases: including or not including a CS fluid in universe.

\section{$\text{Behaviors  of $G$ with the confidence level  in VG-UM theory with or without a CS fluid}$}

\begin{figure}[ht]
   \includegraphics[width=4.6cm]{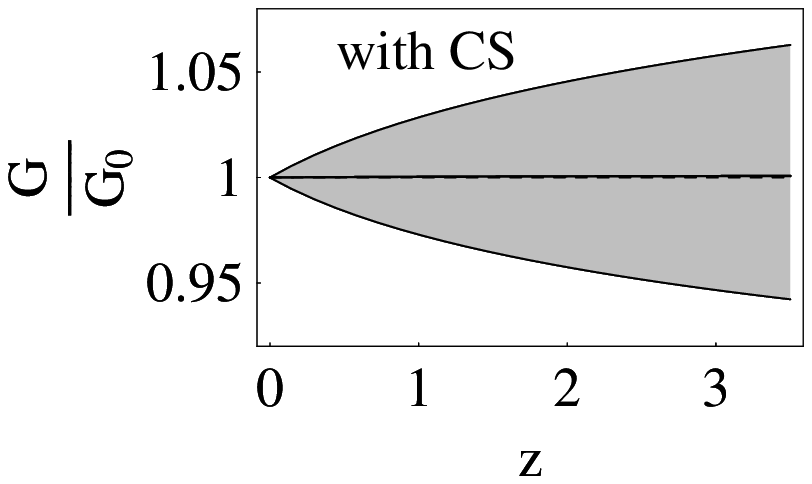}
   \includegraphics[width=4.8cm]{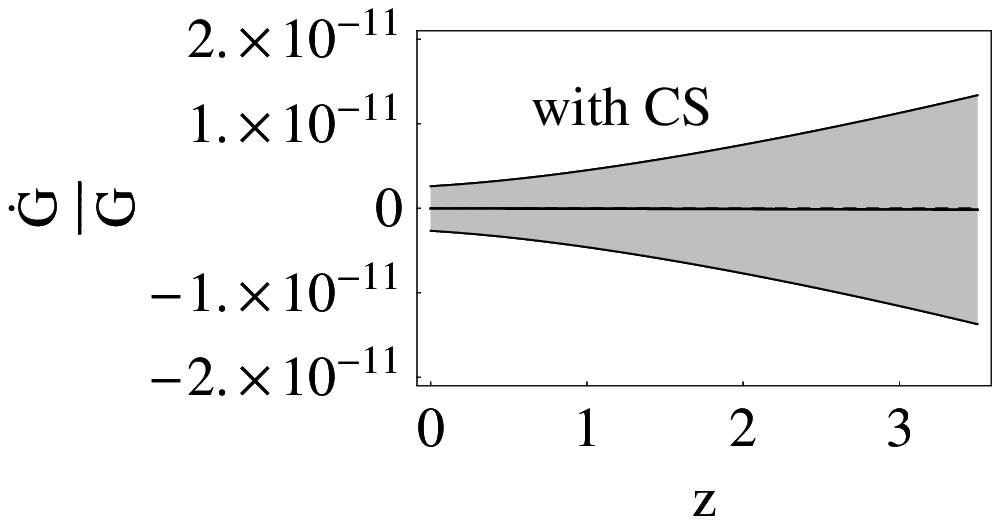}\\
  \includegraphics[width=4.6cm]{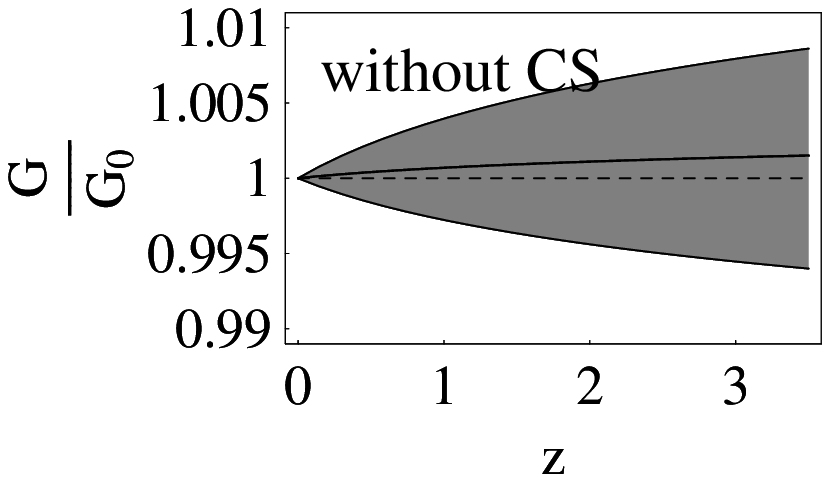}
  \includegraphics[width=4.8cm]{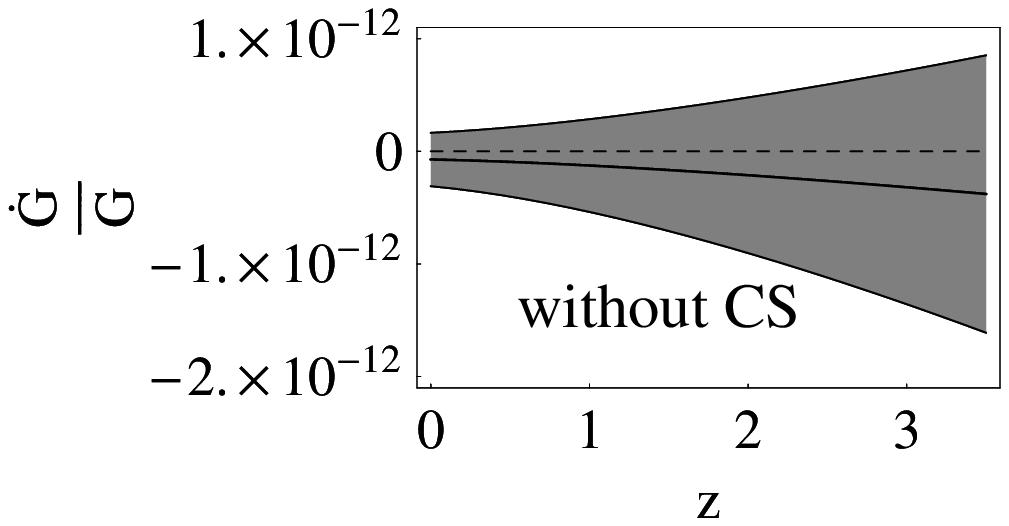}
  \caption{The best-fit evolutions of $\frac{{G}}{G_{0}}$ and $\frac{\dot{G}}{G}$ with their  confidence level in  VG-UM model containing  (or not containing) a CS fluid.}\label{figure-G-vgum0-sigma}
\end{figure}

\begin{table}[!htbp]
 \vspace*{-12pt}
 \begin{center}
 \begin{tabular}{ |c|| c| c | } \hline\hline
         & With CS      &Without CS     \\\hline
   $(\frac{G}{G_{0}})_{z=3.5}$
   & $1.0008^{+0.0620}_{-0.0584}$&$1.0015^{+0.0071}_{-0.0075}$ \\\hline
   $(\frac{\dot{G}}{G})_{today}$
  & $-0.3496^{+26.3135}_{-26.3135}\times 10^{-13} yr^{-1}$
    & $-0.7252^{+2.3645}_{-2.3645}\times 10^{-13} yr^{-1}$    \\\hline
   $(\frac{\dot{G}}{G})_{z=3.5}$
    & $-1.800^{+135.396}_{-135.396}\times 10^{-13} yr^{-1}$
      &$-3.792^{+12.314}_{-12.314}\times 10^{-13} yr^{-1}$ \\\hline
  \hline
 \end{tabular}
 \end{center}
 \caption{The best-fit values of $\frac{G}{G_{0}}$ and $\frac{\dot{G}}{G}$ with their confidence level in VG-UM model containing (or not containing) a CS  fluid. }\label{table-G-vgum0}
 \end{table}

\begin{table}
\begin{tabular}{|c|c|}\hline\hline
   Observations                                                &  Limits $(yr^{-1})$\\\hline
   Pulsating white dwarf G117-B15A \cite{VG-MNRAS-2004-dwarf}  & $\mid \frac{\dot{G}}{G}\mid  \leq 4.1 \times 10^{-10} $ \\\hline
   Nonradial pulsations of white dwarfs \cite{VG-PRD-2004-white} & $-2.5 \times 10^{-10}\leq\frac{\dot{G}}{G}\leq 4\times 10^{-11} $\\\hline
   Millisecond pulsar PSR J0437-4715 \cite{VG-APJ}             & $\mid \frac{\dot{G}}{G}\mid \leq   2.3 \times 10^{-11}$ \\\hline
   Type-Ia Supernovae \cite{VG-PRD-2002-SN}                    & $ \frac{\dot{G}}{G}\leq 10^{-11} $\\\hline
   Neutron star masses \cite{VG-PRL-1996-neutron}              & $ \frac{\dot{G}}{G}= (-0.6\pm 4.2) \times 10^{-12} $ \\\hline
   Helioseismology  \cite{VG-APJ-1998}                         & $\mid \frac{\dot{G}}{G}\mid \leq   1.6 \times 10^{-12} $ \\\hline
   Lunar laser ranging experiment \cite{VG-PRL-2004-LL}        & $ \frac{\dot{G}}{G}= (4\pm 9) \times 10^{-13} $ \\\hline
   Big Bang Nuclei-synthesis \cite{VG-PRL-2004-BB}             & $ -3.0 \times 10^{-13}<\frac{\dot{G}}{G}<4.0\times 10^{-13} $\\\hline \hline
\end{tabular}
\caption{Limits  on the variation of $G$.}\label{table-G-observations}
\end{table}

 In VG-UM theory with or without  a CS fluid,  the best-fit evolutions of  $\frac{\dot{G}}{G}$ with their confidence level  (the shadow region) are  illustrated in  Figure \ref{figure-G-vgum0-sigma} by using the best-fit values of model parameters with their covariance matrix.  "Dot" denotes the derivative with respect to $t$. In the VG-UM model with a CS fluid, limit on the variation of $G$ at today is $(\frac{\dot{G}}{G})_{today}\simeq -0.3496^{+26.3135}_{-26.3135}\times 10^{-13}yr^{-1}$, and at $z= 3.5$ we have $(\frac{G}{G_{0}})_{z=3.5}\simeq 0.9917^{+0.0104}_{-0.0131}$ and  $(\frac{\dot{G}}{G})_{z=3.5}\simeq -1.800^{+135.396}_{-135.396} \times 10^{-13} yr^{-1}$.   For case without a CS fluid, Fig. \ref{figure-G-vgum0-sigma}  shows the prediction that the today's value is $(\frac{\dot{G}}{G})_{today}\simeq -0.7252^{+2.3645}_{-2.3645}\times 10^{-13} yr^{-1}$. This restriction on  $(\frac{\dot{G}}{G})_{today}$ is more stringent   than other results seen in table \ref{table-G-observations}.  Also, using the best-fit value of parameter $\beta$  with error the shapes of $\frac{G}{G_{0}}=(1+z)^{\beta}$ are exhibited. Taking high redshift $z= 3.5$ as another reference points,   we find $(\frac{G}{G_{0}})_{z=3.5}\simeq 1.0015^{+0.0071}_{-0.0075}$ and  $(\frac{\dot{G}}{G})_{z=3.5}\simeq    -0.3792^{+1.2314}_{-1.2314} \times 10^{-12} yr^{-1}$ in the  VG-UM model without a CS fluid.   It is important to sternly constrain the value of $\beta$, since the  monotonicity of $\frac{\dot{G}}{G}=-\beta H$ depends on the symbol of $\beta$.  Fig. \ref{figure-G-vgum0-sigma} reveal that the behaviors of $G$ and its derivative  are around the constant-$G$ theory  for both cases: including or not including a CS fluid in universe.

\section{$\text{Behaviors  of EoS  with the confidence level  in VG-UM theory with or without  a CS fluid}$}

  The EoS of   UM in VG theory is demonstrated
\begin{eqnarray}
w_{VG-UM}(z)=\frac{p_{VG-UM}}{\rho_{VG-UM}}=\frac{\beta-3}{3}\frac{B_{s}}{B_{s}+(1-B_{s})(1+z)^{(1+\alpha)(3-\beta)}}.\label{w-vgum0}
\end{eqnarray}
 From Figure  \ref{figure-w-vgum0-sigma} (left), we can see that $w_{VG-UM}\sim0$ (DM) at early time and $w_{VG-UM}\sim -1$ (DE)  in the future for the VG-UM model with  or  without a CS fluid. If  the dark sectors are thought to be separable, it is interested to investigate  the properties of both dark components in VG-UM  model. Supposing that the behavior of dark matter is known i.e. its EoS  $w_{dm}=0$ ($\rho_{dm}=\rho_{0dm}a^{\frac{-\beta^{2}-2\beta-6}{2+\beta}}$), the EoS of  dark energy in  VG-UM model subjects to
\begin{eqnarray}
w_{de}=\frac{p_{de}}{\rho_{de}}=\frac{p_{VG-UM}}{\rho_{VG-UM}-\rho_{dm}}=\frac{-A}{\rho_{VG-UM}^{1+\alpha}-\rho_{dm}\rho_{VG-UM}^{\alpha}}.
\end{eqnarray}
Using the best-fit values of model parameters and the covariance matrix, the evolutions of $w_{de}$ with confidence level in VG-UM model containing (or not containing) a CS fluid are plotted in Fig. \ref{figure-w-vgum0-sigma} (middle). If one deems the behavior of dark energy is the cosmological constant i.e. $w_{\Lambda}=-1$ ($p_{\Lambda}=-\rho_{\Lambda}$), the EoS of dark matter in VG-UM model obeys
\begin{eqnarray}
w_{dm}=\frac{p_{dm}}{\rho_{dm}}=\frac{p_{VG-UM}-p_{\Lambda}}{\rho_{VG-UM}-\rho_{\Lambda}}
=\frac{\rho_{\Lambda}\rho_{VG-UM}^{\alpha}-A}{\rho_{VG-UM}^{1+\alpha}-\rho_{\Lambda}\rho_{VG-UM}^{\alpha}}\label{w-dm}
\end{eqnarray}
which is drawn in Fig. \ref{figure-w-vgum0-sigma} (right)  with the confidence level for two cases (with or without a  CS fluid).

\begin{figure}[ht]
  \includegraphics[width=4.6cm]{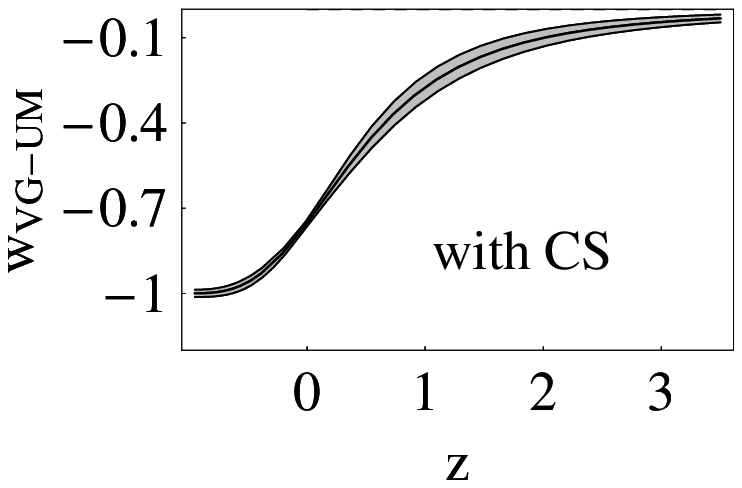}
  \includegraphics[width=4.6cm]{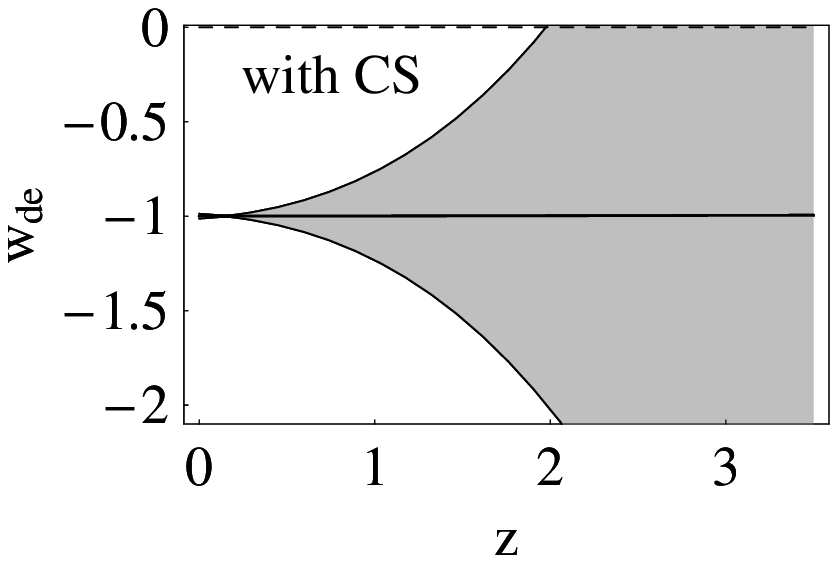}
  \includegraphics[width=4.6cm]{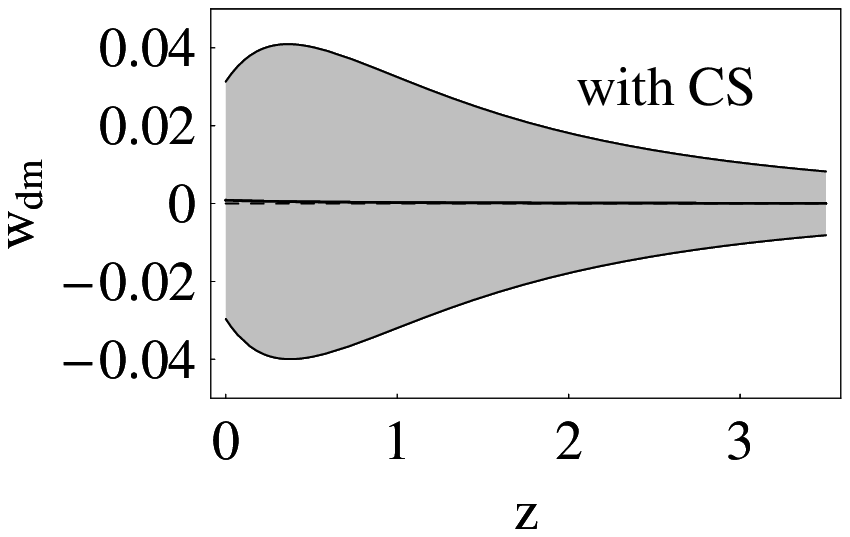}\\
  \includegraphics[width=4.6cm]{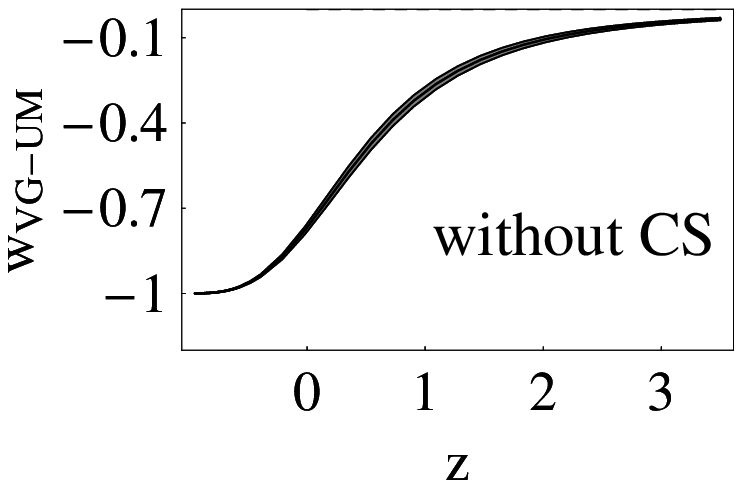}
  \includegraphics[width=4.6cm]{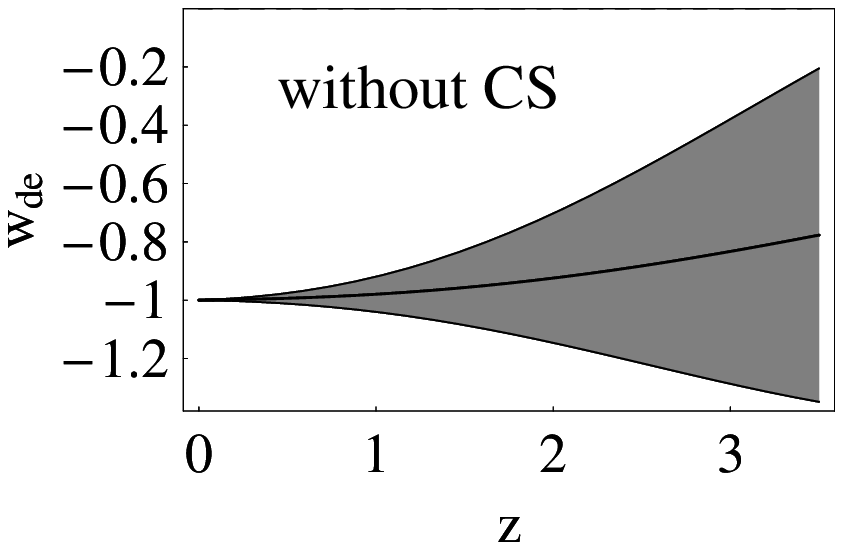}
  \includegraphics[width=4.6cm]{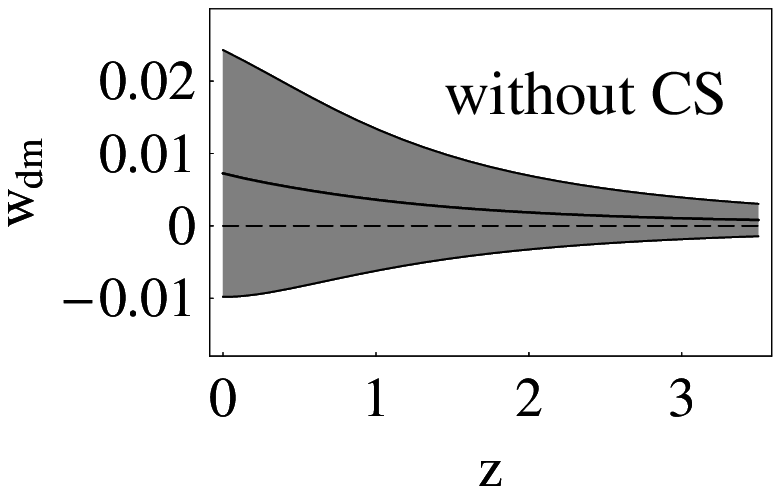}\\
  \caption{The evolutions of EoS with confidence level in VG-UM model including   (lower)  or not including (upper) a CS fluid.  Evolution of  $w_{VG-UM}(z)$  (left),  evolution of $w_{de}(z)$  in VG-UM model with assuming $w_{dm}=0$ at prior  (middle),  and  evolution of $w_{dm}(z)$  in VG-UM model  with assuming $w_{de}=-1$ at prior (right).}\label{figure-w-vgum0-sigma}
\end{figure}

\begin{table}[!htbp]
 \vspace*{-12pt}
 \begin{center}
 \begin{tabular}{ |c|| c| c| c | } \hline\hline
          & $w_{0VG-UM}$ &   $w_{0dm}$ (with $w_{0de}=-1$)    &  $w_{0de}$  (with $w_{0dm}=0$)       \\\hline
 With CS  & $-0.7519^{+0.0112}_{-0.0112}$  & $0.0009^{+0.0304}_{-0.0304}$    & $-0.9998^{+0.0125}_{-0.0125}$  \\\hline
  Without CS     & $-0.7438^{+0.0134}_{-0.0134}$  & $0.0072^{+0.0170}_{-0.0170}$    & $-0.9986^{+0.0011}_{-0.0011}$   \\\hline\hline
 \end{tabular}
 \end{center}
 \caption{The best-fit values of  $w_{0VG-UM}$, $w_{0dm}$ and $w_{0de}$ with their confidence level hinted by VG-UM model with or without a CS  fluid. }\label{table-w-vgum0}
 \end{table}

  From Fig. \ref{figure-w-vgum0-sigma}, we get the current values $w_{0dm}=0.0009^{+0.0304}_{-0.0304}$ in  VG-UM model with a CS fluid and $w_{0dm}=0.0072^{+0.0170}_{-0.0170}$  in  VG-UM model without a CS fluid, which have the larger uncertainties than $w_{0dm}=0.0010^{+0.0016}_{-0.0016}$ calculated on the non-unified model of constant-$G$ theory by Ref. \cite{wdm}. For the current value $w_{0de}$, it approximates to -1  with the very small uncertainty for both    VG-UM model with a CS fluid ($w_{0de}=-0.9998^{+0.0125}_{-0.0125}$) and VG-UM model without a CS fluid ($w_{0de}=-0.9986^{+0.0011}_{-0.0011}$). From the best fit evolution  in VG-UM model with a CS fluid, we can see that both $w_{de} (\sim -1)$ and $w_{dm} (\sim 0)$ tends to be constant, but  the uncertainties of them are much larger than that in model without a CS fluid.  For the best-fit evolution in VG-UM model without a CS fluid,  $w_{de}$ and $w_{dm}$ are variable with the time and $w_{dm}$  tends to have  small deviation from zero (small-positive pressure) at the recent time. In addition, at high redshift  the uncertainty of $w_{de}$ (or $w_{dm}$) is enlarged (or narrowed) for both VG-UM model with a CS fluid  and VG-UM model without a CS fluid.

\section{$\text{Perturbational behaviors in structure formation for VG-UM theory}$}

The study on the structure formation is necessary for a cosmological theory. We investigate the evolutions of growth function $f$ and growth factor $D$ in VG-UM theory. The derivation of evolutionary equation $f$ and $D$ are shown in appendix. Using the definition $f(a)=\frac{aD^{'}(a)}{D(a)}=\frac{d\ln \delta}{d\ln a}$, we obtain the dynamically evolutionary equation of $f$
\begin{equation}
(1+z)f'-f^{2}+(1+z)f\frac{E'}{E}-\frac{4+2\beta+2\beta^{2}}{2+\beta}f+\frac{6+2\beta
+\beta^{2}}{(2+\beta)^{2}}\frac{\Omega_{0m}}{E^{2}}(1+z)^{\frac{6+2\beta+\beta^{2}}{2+\beta}}=0\label{vg-fz-eq}
\end{equation}
where prime denotes the derivative with respect to redshift $z$ and $E(z) = H(z)/H_{0}$.

\begin{figure}[ht]
  \includegraphics[width=5.5cm]{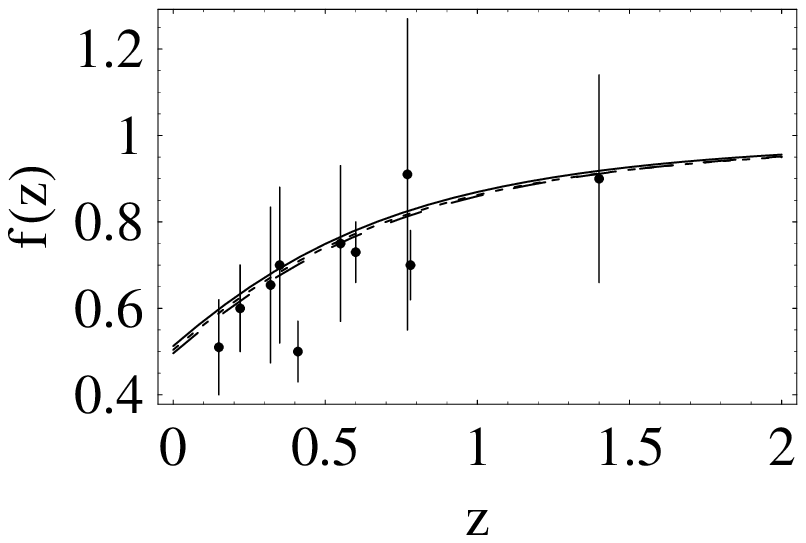}
  \includegraphics[width=5.5cm]{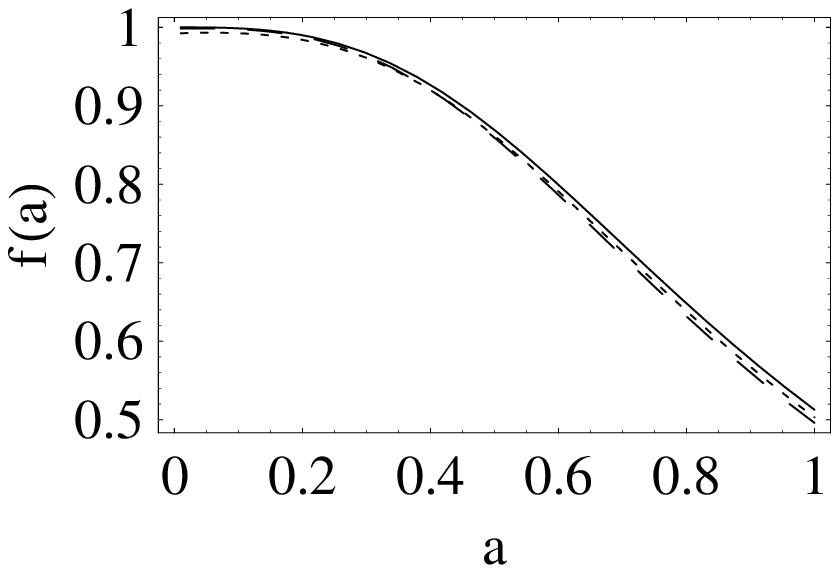}
  \includegraphics[width=5.5cm]{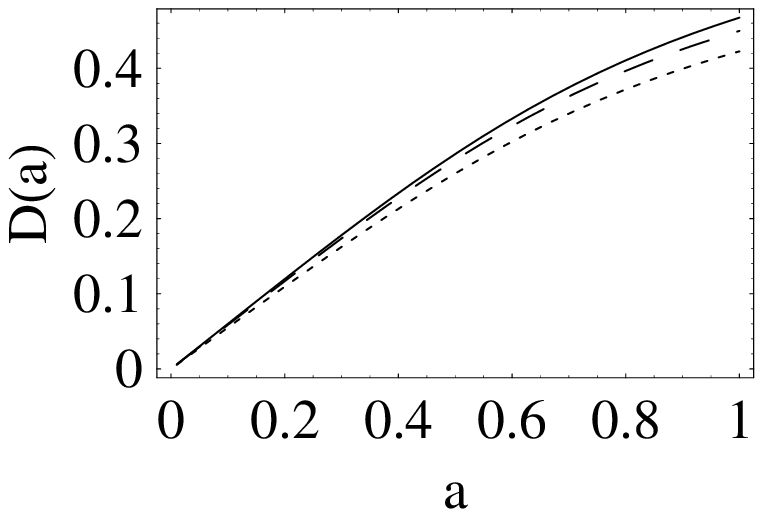}\\
  \caption{The evolutions of $f(z)$, $f(a)$ and $D(a)$. The solid lines correspond to the $\Lambda$CDM model,  the short-dash lines correspond to the VG-UM model with a CS fluid, and the dot lines correspond to the VG-UM model without a CS fluid.}\label{figure-D-f}
\end{figure}
In Fig. \ref{figure-D-f}, we use the best-fit  values of cosmological parameters in table \ref{table-vgum0-with} and \ref{table-vgum0-without} to plot the evolutions of growth function $f $ and growth factor $D$ for VG-UM model and $\Lambda$CDM model by numerically solving Eq. (\ref{D-evolutoin-text}) and (\ref{vg-fz-eq})  with the  initial conditions $a_{i} = 0.0001$, $D(a_{i}) = a_{i}$, $D^{'}(a_{i}) = 0$ and $f(a_{i})=1$. We can see that the evolutions of $f(a)$ for VG-UM model (including or not including a CS fluid) fit  well as $\Lambda$CDM model, and the behavior of $f(z)$ are well consistent with the observational growth data listed in table \ref{table-f-data}. In  VG-UM model with or without a CS fluid, $D(a)$ evolves slower  (more slow  growth of perturbations) than that in the $\Lambda$CDM model. The current value of $D(a = 1)$ in $\Lambda$CDM model  is approximately 12\% larger than that in VG-UM model without a CS fluid.

\section{$\text{Conclusions}$}
 Observations anticipate that $G$ may be variable and most universal energy density are invisible.  The attractive properties of this study is that  the variation of $G$ naturally results to the invisible components in universe. The VG  could provide a solution to the originated  problem of DM and DE. We apply recently  observed data  to constrain the unified model of dark sectors   with or without a CS fluid in the framework of  VG theory. Using the LT, the GRBs, the GF, the SNIa with systematic error, the CMB from 9-year WMAP and the BAO data from measurement of radial and peak positions, uncertainties of VG-UM parameter space are obtained.

 For the case without a cosmic string fluid, constraint on mean value of VG parameter is $\beta=0.0007^{+0.0032+0.0062}_{-0.0033-0.0067}$ with a small uncertainty around zero, and  restrictions  on UM model parameters are $B_{s}=0.7442^{+0.0137+0.0262}_{-0.0132-0.0292}$ and $\alpha=0.0002^{+0.0206+0.0441}_{-0.0209-0.0422}$ with  $1\sigma$ and  $2\sigma$ confidence level. For the case with a cosmic string fluid, restriction on dimensionless  density parameter of CS fluid  is  $\Omega_{s}=-0.0106^{+0.0312+0.0582}_{-0.0305-0.0509}$ in the VG-UM theory. Obviously,  the uncertainty of $\Omega_{s}$ is larger than some results on $\Omega_{k}$ in the framework of $G$-constant theory. At $1\sigma$ confidence level the flat $\Lambda$CDM model ($\Omega_{s}=0$, $\beta=0$ and $\alpha=0$) is included in the VG-UM model.

 Using the best-fit values of VG-UM parameters and their covariance matrix, the limits on today's value are $(\frac{\dot{G}}{G})_{today}= -0.7252^{+2.3645}_{-2.3645} \times 10^{-13}$ or  $(\frac{\dot{G}}{G})_{today}\simeq -0.3496^{+26.3135}_{-26.3135}\times 10^{-13}yr^{-1}$ for the universe with or without a CS fluid. And corresponding to these two cases, we finds  $(\frac{G}{G_{0}})_{z=3.5}\simeq 1.0015^{+0.0071}_{-0.0075}$ and  $(\frac{G}{G_{0}})_{z=3.5}\simeq 1.0008^{+0.0620}_{-0.0584}$ at redshift $z=3.5$. If one considers that the DM and the DE could be  separable in unified model, EoS of DE and DM are discussed by combing with the fitting results. It is shown that $w_{0dm}=0.0072^{+0.0170}_{-0.0170}$ or $w_{0dm}=0.0009^{+0.0304}_{-0.0304}$   with assuming $w_{0de}=-1$ for  VG-UM universe containing or not containing a CS fluid, while there are $w_{0de}=-0.9986^{+0.0011}_{-0.0011}$ or  $w_{0de}=-0.9998^{+0.0125}_{-0.0125}$  with assuming $w_{0dm}=0$ at prior for VG-UM model with or without  a CS fluid.

 \textbf{\ Acknowledgments }
 The research work is supported by   the National Natural Science Foundation of China (11205078,11275035,11175077).

\appendix

\section{$\text{The growth of structures in linear perturbation theory}$}

In a sub-horizon region with length scale $r<H^{-1}$, the density of DE and  cold DM  are expressed by $\tilde{\rho}_{sde}$ and $\tilde{\rho}_{sdm}$, respectively. We suppose that DE is not perturbed,  DM is perturbed in sub-horizon region. So, we have $\tilde{\rho}_{sde}=\rho_{de}$ for  the homogeneous DE in whole universe and $\tilde{\rho}_{sdm}=\rho_{dm}+\delta\rho_{dm}$  for the perturbed DM, where  $\rho_{de}$ and $\rho_{dm}$ denote the density of  DE and DM in background level, respectively. Obviously, the region of  $\delta\rho_{dm}>0$ will cluster and form structure. In analogy to the equation in background level, the evolution of matter density inside the perturbed region can be given by the following conservation equation
\begin{equation}
\dot{\tilde{\rho}}_{sdm}+3h(\frac{6+2\beta+\beta^2}{6+3\beta}\tilde{\rho}_{sdm}+\frac{2+2\beta}{2+\beta}\tilde{p}_{sdm})=0.\label{rhomc-cont-inside}
\end{equation}
Symbol "tilde" denotes the cosmological quantity in perturbed region. In this region, the local expansion is described by $h = \dot{r}/r$ and the acceleration is
\begin{equation}
\frac{\ddot{r}}{r}=-\frac{8\pi G(t)}{3(2+\beta)}(\tilde{\rho}_{sdm}+\rho_{de}+3\tilde{p}_{sdm}+3p_{de})-\frac{\beta^{2}}{(2+\beta)}h\label{rr-inside}
\end{equation}
which is same as Eq.(\ref{acc-vg}) for background level. One can define the density contrast of DM
\begin{equation}
1+\delta_{dm}=\frac{\tilde{\rho}_{sdm}}{\rho_{dm}}\label{contrast}
\end{equation}
with $\delta_{dm}>0$. Differentiating Eq. (\ref{contrast}) with respect to $t$ gives
\begin{equation}
\dot{\delta}_{dm}+(\frac{6+2\beta+\beta^2}{2+\beta})(1+\delta_{dm})(h-H)+\frac{6+6\beta}{2+\beta}(1+\delta_{dm})(h\tilde{w}_{sdm}-Hw_{dm})=0\label{dotdeta}
\end{equation}
after using Eqs.  (\ref{rhomc-cont-inside}) and (\ref{conservation}). Taking the time derivative in above equation obtains
\begin{equation}
\ddot{\delta}_{dm}-\frac{\dot{\delta}_{dm}^2}{1+\delta_{dm}}+(\frac{6+2\beta+\beta^2}{2+\beta})(1+\delta_{dm})(\dot{h}-\dot{H})+
\frac{6+6\beta}{2+\beta}(\dot{h}\tilde{w}_{sdm}+h\dot{\tilde{w}}_{sdm}-\dot{H}\tilde{w}_{sdm}-H\dot{\tilde{w}}_{sdm})(1+\delta_{dm})=0,\label{ddotdeta}
\end{equation}
 where
\begin{equation}
\dot{h}-\dot{H}=-\frac{H^2}{2+\beta}\Omega_{dm}\delta_{dm}-\frac{4+2\beta+2\beta^2}{2+\beta}(h-H)H-\frac{3H^2}{2+\beta}(\frac{\tilde{\rho}_{sdm}}{\tilde{\rho}_{c}}\tilde{w}_{sdm}
-\Omega_{dm}w_{dm})\label{dh}
\end{equation}
is given by substituting Eqs. (\ref{acc-vg}) and (\ref{rr-inside}) into $\dot{H}=\frac{\ddot{a}}{a}-H^2$ and $\dot{h}=\frac{\ddot{r}}{r}-h^2$, respectively. In addition, in calculation we used $\rho_c=3H^2/8\pi G(t)$ and $h+H\simeq 2H$.  Inserting (\ref{dh}) into (\ref{ddotdeta}) results
\begin{eqnarray}
\ddot{\delta}_{dm}-\frac{\dot{\delta}_{dm}^2}{1+\delta_{dm}}+\frac{4+2\beta+\beta^{2}}{2+\beta}H\dot{\delta}_{dm}-
(\frac{6+2\beta+\beta^2}{2+\beta})\frac{H^2\Omega_{dm}}{2+\beta}(\delta_{dm}+\delta_{dm}^{2}) \nonumber\\
+[\frac{4+2\beta+2\beta^2}{2+\beta}\frac{6+6\beta}{2+\beta}H(h\tilde{w}_{sdm}-Hw_{dm})
+\frac{6+6\beta}{2+\beta}(\dot{h}w_{dm}+h\dot{\tilde{w}}_{sdm}-\dot{H}w_{dm}-H\dot{w}_{dm}) \nonumber\\
+(\frac{6+2\beta+\beta^2}{2+\beta})\frac{3H^2}{2+\beta}(\frac{\tilde{\rho}_{sdm}}{\tilde{\rho}_{c}}\tilde{w}_{sdm}-\Omega_{dm}w_{dm})](1+\delta_{dm})=0.\label{ddotdeta2}
\end{eqnarray}
 Neglecting square terms of $\delta_m$ in (\ref{ddotdeta2}), we receive the evolutional equation of density contrast in  spherical overdense region
 \begin{eqnarray}
\ddot{\delta}_{dm}+\frac{4+2\beta+2\beta^{2}}{2+\beta}H\dot{\delta}_{dm} -(\frac{6+2\beta +\beta^2}{2+\beta})\frac{H^2\Omega_{dm}}{2+\beta}\delta_{dm}
+[\frac{4+2\beta+2\beta^2}{2+\beta}\frac{6+6\beta}{2+\beta}H(h\tilde{w}_{sdm}-Hw_{dm}) \nonumber\\
+\frac{6+6\beta}{2+\beta}(\dot{h}\tilde{w}_{sdm}+h\dot{w}_{dm}-\dot{H}\tilde{w}_{sdm}-H\dot{w}_{sdm})
+(\frac{6+2\beta+\beta^2}{2+\beta})\frac{3H^2}{2+\beta}(\frac{\tilde{\rho}_{sdm}}{\tilde{\rho}_{c}}\tilde{w}_{sdm}-\Omega_{dm}w_{dm})](1+\delta_{dm}) =0.\label{ddotdeta3}
\end{eqnarray}
 Taking $\beta=0$, the above equation reduces to case of constant $G$ given by reference \cite{D-G}. Using the definition of growth factor $D(a)$,  we can rewrite Eq. (\ref{ddotdeta3})  as follows
\begin{eqnarray}
 D^{''}(a)+[\frac{E^{'}(a)}{E(a)}+\frac{1}{a}+\frac{4+2\beta+2\beta^{2}}{a(2+\beta)}]D^{'}(a)- \frac{6+2\beta+\beta^{2}}{(2+\beta)^{2}}\frac{\Omega_{0dm}}{E(a)^{2}a^{2}}a^{\frac{-6-2\beta-\beta^{2}}{2+\beta}}D(a) \nonumber\\
 +[\frac{(4+2\beta+2\beta^2)(6+6\beta)}{(2+\beta)^{2}a^{2}H}(h\tilde{w}_{sdm}-Hw_{dm})
 +\frac{6+6\beta}{(2+\beta)aH}(h^{'}\tilde{w}_{sdm}+h\tilde{w}^{'}_{sdm}-H^{'}w_{dm}-Hw^{'}_{dm})  \nonumber\\
  +\frac{18+6\beta+3\beta^2}{(2+\beta)^{2}a^{2}}(\frac{\tilde{\rho}_{sdm}}{\tilde{\rho}_{c}}\tilde{w}_{sdm}-\Omega_{dm}w_{dm})][1+D(a)\delta_{dm}(a=1)]=0.\label{D-evolutoin}
 \end{eqnarray}
 The linear regime of cosmological perturbations is valid for all sales during the early radiation dominated era and for most sales during the matter dominated era.  For $w_{dm}\simeq w_{sdm}\simeq 0$, above equation reduces to
\begin{eqnarray}
 D^{''}(a)+[\frac{E^{'}(a)}{E(a)}+\frac{1}{a}+\frac{4+2\beta+2\beta^{2}}{a(2+\beta)}]D^{'}(a)- \frac{6+2\beta+\beta^{2}}{(2+\beta)^{2}}\frac{\Omega_{0dm}}{E(a)^{2}a^{2}}a^{\frac{-6-2\beta-\beta^{2}}{2+\beta}}D(a)=0.\label{D-evolutoin1}
 \end{eqnarray}
Transferring the function from $D$ to $f$ in above equation, we get
\begin{eqnarray}
 f^{'}(a)+\frac{f^{2}(a)}{a}+[\frac{E^{'}(a)}{E(a)}+\frac{2}{a}(\frac{2+\beta+\beta^{2}}{(2+\beta)})]f(a)- \frac{6+2\beta+\beta^{2}}{(2+\beta)^{2}}\frac{\Omega_{0dm}}{E(a)^{2}a}a^{\frac{-6-2\beta-\beta^{2}}{2+\beta}}=0.\label{f-evolutoin1}
 \end{eqnarray}

\end{document}